\documentclass[reprint,amsmath,amssymb,aps,superscriptaddress,pre]{revtex4-2}
\usepackage{graphicx}
\usepackage{color}
\usepackage{hyperref}
\hypersetup{
    colorlinks=true,
    linkcolor=blue,
    citecolor=blue,
    urlcolor=black,
}
\usepackage{comment}
\usepackage{etoolbox}
\usepackage{caption}
\usepackage{booktabs}
\usepackage{siunitx}
\usepackage{bm}
\usepackage{gensymb}

\newcommand{\add}[1]{\textcolor{black}{#1}}

\begin{document}

\title{A Local Structural Basis to Resolve Amorphous Ices}

\author{Quinn M. Gallagher}
\thanks{These authors contributed equally to this work.}
\affiliation{Department of Chemical and Biological Engineering, Princeton University, Princeton, NJ 08540, United States}

\author{Ryan J. Szukalo$^{*}$}
\email{rszukalo@princeton.edu}
\affiliation{Department of Chemistry, Princeton University, Princeton, NJ 08540, United States}

\author{Nicolas Giovambattista}
\affiliation{Department of Physics, Brooklyn College of the City University of New York, Brooklyn, New York 11210, United States}
\affiliation{Ph.D. Programs in Physics and Chemistry, The Graduate Center of the City University of New York, New York, New York 10016, United States}

\author{Pablo G. Debenedetti}
\affiliation{Department of Chemical and Biological Engineering, Princeton University, Princeton, NJ 08540, United States}

\author{Michael A. Webb}
\email{mawebb@princeton.edu}
\affiliation{Department of Chemical and Biological Engineering, Princeton University, Princeton, NJ 08540, United States}

\begin{abstract}
Phases with distinct thermodynamic properties must differ in their underlying microscopic configurations. 
While ordered phases are readily distinguished by unit cells and space groups, the local structural basis differentiating amorphous phases is less apparent. 
Here, using a new probabilistic data-driven framework applied to molecular simulations of water, we identify local collective variables that discriminate low-density and high-density amorphous (LDA and HDA) ices and characterize pressure-induced transitions between them. 
As expected, descriptors related to local density effectively distinguish LDA and HDA; however, phase identity is surprisingly encoded within the first coordination shell.
Furthermore, the pressure-induced LDA--HDA transformation proceeds through redistribution between LDA- and HDA-like local environments with no evidence for intermediate structures, consistent with a first-order-like phase transition. 
This contrasts with the gradual structural evolution observed in other amorphous systems, such as metallic glasses.
Critically, local hydrogen density reveals pronounced structural hysteresis between compression and decompression pathways, which is not apparent in orientational order parameters, demonstrating that the microscopic interpretation of amorphous transformations depends fundamentally on descriptor choice.
These findings are robust across force fields and provide a general strategy for characterizing disordered phases lacking obvious distinguishing features.
\end{abstract}

\maketitle

The thermodynamic properties that define macroscopic materials ultimately emerge from distributions of microscopic configurations \cite{Landau:1980}.
In crystalline materials, the structural basis of a particular phase is unambiguously in its unit cell and space group.
For amorphous phases, there is no obvious analogous structural scaffold. 
If two amorphous phases are thermodynamically distinct, with different densities or response functions, their underlying distributions of local environments must also differ \cite{Debenedetti:2001, Berthier:2011}.
What is far less clear is how they differ, over what length scales such differences are encoded, and whether they can be expressed in terms of a small number of collective variables that admit a microscopic interpretation.

Water's amorphous ices provide a compelling arena to address these questions.
Low-density amorphous (LDA) and high-density amorphous (HDA) ice represent two glassy states with markedly different densities and thermodynamic behavior \cite{Mishima:1984, Mishima:1985}. 
Their interconversion has long been discussed in the context of water's putative liquid--liquid phase transition and associated polyamorphism~\cite{Poole:1992, Gallo:2016, Debenedetti:2020}.
LDA and HDA are viewed as glassy analogs of the low-density and high-density liquids in supercooled water~\cite{Amann:2013, Kim:2020}, making the structural basis directly relevant to long-standing questions about water's anomalous behavior.

\begin{figure*}
    \centering
    \includegraphics[width=0.9\textwidth]{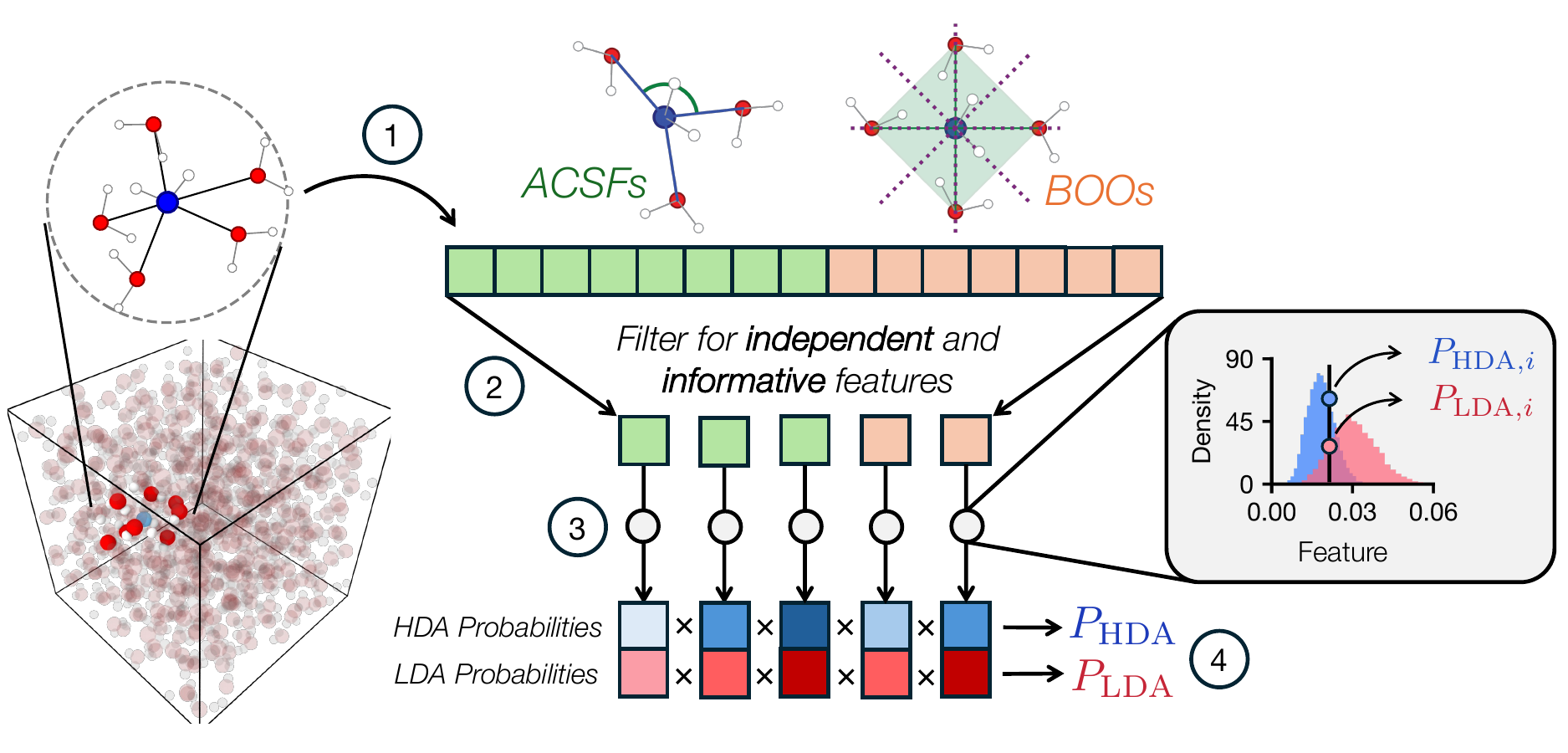}
    \captionsetup{justification=raggedright, font=small}
    \caption{
    \textbf{Probabilistic framework for resolving local structural differences between amorphous phases.}
    (1) Local atomic environments are sampled from MD simulations and represented using ACSF and BOO descriptors.  
    (2) A two-stage feature-selection procedure removes redundant descriptors and identifies those most informative for distinguishing LDA and HDA.  
    (3) For each selected descriptor and class, one-dimensional probability densities are estimated using Gaussian kernel density estimation.  
    (4) The joint probability for each class is computed as the product of the descriptor-wise probabilities, providing an interpretable probabilistic classification and enabling explicit detection of out-of-distribution configurations.}
    \label{fig:explanation}
\end{figure*}

Prior studies have established that LDA exhibits an open, tetrahedrally ordered network, whereas HDA is denser and more disordered~\cite{Finney:2002, Loerting:2006, Gallo:2016}.
This picture has emerged primarily from physically motivated descriptors, including tetrahedral order parameters~\cite{Chau:1998, Errington:2001, Jabes:2010}, bond-orientational order (BOO) metrics~\cite{Steinhardt:1983, Martelli:2020, Faure:2024}, coordination numbers, and local structure indices~\cite{Errington:2001, Foffi:2022, Donkor:2024}.
However, which descriptors most effectively encode this distinction, and whether other key collective variables remain undiscovered or more informative, is unclear.
Studies of supercooled liquid water suggest that distinguishing low- and high-density environments requires correlations extending beyond the first coordination shell~\cite{Martelli:2020, Donkor:2024}, but whether this nonlocality reflects fundamental structural physics rather than descriptor limitations has not been assessed.

A separate open question concerns the microscopic character of the LDA--HDA transformation.
These glasses exhibit first-order-like transformations under pressure~\cite{Mishima:1984, Klotz:2005, Loerting:2006, Chiu:2013, Wong:2015, Szukalo:2025}, yet whether the transition proceeds through structurally distinct intermediates, as in some metallic glasses~\cite{Zeng:2010, Du:2019}, or through redistribution between LDA-like and HDA-like motifs without intermediates~\cite{Martelli:2020, Dhabal:2023, Ramesh:2024}, remains unresolved at the molecular scale.

Existing approaches struggle to address these questions simultaneously.
Simple low-dimensional order parameters are interpretable but can be biased by their construction, highlighting anticipated features while obscuring others.
High-dimensional descriptors~\cite{Behler:2011, Bartok:2013} can encode rich local information, and machine-learning classifiers built on these representations can achieve high discriminative accuracy~\cite{Defever:2019, Schmidt:2019, Ceriotti:2019, Musil:2021, Boattini:2019, Faure:2024, Cai:2025}.
However, such models are often opaque, classifying without revealing which descriptors matter and lacking mechanisms to detect out-of-distribution (OOD) configurations. 
One therefore cannot identify the minimal discriminative features, assess how local the distinction is, or determine whether transformation intermediates represent novel structures.

Here, we use an interpretable probabilistic framework applied to molecular simulations of water (Figure~\ref{fig:explanation}) to systematically identify the local structural basis distinguishing LDA and HDA and to characterize the microscopic nature of their pressure-induced interconversion.
Mutual-information (MI) analysis ranks a large set of local structural descriptors by their discriminative power, and the most informative are combined into a calibrated classifier that also detects configurations belonging to neither phase.
We find that interstitial hydrogen density dominates the discrimination and that phase identity is encoded within the first coordination shell. 
Analysis of transformation trajectories reveals no intermediate structures; rather, the transition proceeds through redistribution of discrete LDA-like and HDA-like environments. 
Strikingly, local hydrogen density exposes structural hysteresis between compression and decompression pathways that is entirely invisible to orientational order parameters, revealing that the choice of descriptor has fundamental consequences for the physical interpretation of amorphous transformations.

\section*{Probabilistic classification effectively distinguishes amorphous states and novel structures}

We consider probabilistic classification of data generated by DP\_MBpol, a Deep Potential model trained on the many-body polarizable (MB-pol) potential parameterized against coupled-cluster calculations~\cite{Bore:2023, Szukalo:2025}.
A second model, DP\_SCAN, trained on DFT data with the SCAN functional~\cite{Sun:2015, Zhang:2021pd}, is analyzed in the SI and employed later to assess sensitivity to force field choice.
Local environments are represented using atom-centered symmetry functions (ACSFs) \cite{Behler:2011} and bond-orientational order (BOO) parameters \cite{Steinhardt:1983}, constructed from 16 nearest neighbors, consistent with prior work \cite{Martelli:2020, Donkor:2024, Faure:2024}, though we systematically vary environment size for analysis.
The framework uses mutual information to rank descriptors by discriminative power, selects an informative and approximately uncorrelated subset, and combines class-conditional density estimates into a joint probability that enables molecule-by-molecule phase assignment and detection of out-of-distribution configurations.
Importantly, this procedure automatically discovers which descriptors best distinguish phases without requiring \textit{a priori} knowledge of the relevant order parameters.
Complete methodological details are provided in Methods and SI.

\begin{figure*}[htb!]
    \centering
    \includegraphics[width=\textwidth]{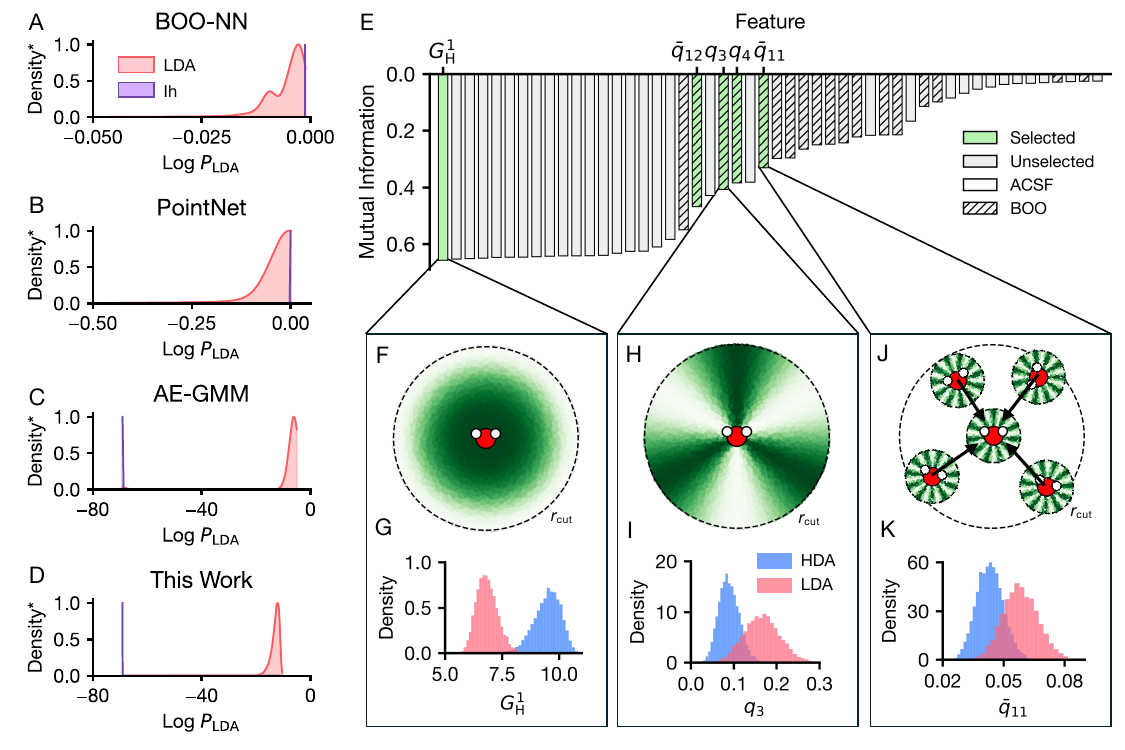}
    \captionsetup{justification=raggedright, singlelinecheck=false,font=small}
    \caption{
    \textbf{Local hydrogen density and orientational symmetry resolve amorphous and crystalline ice environments.} (A-D) Predicted LDA log-probabilities for true LDA environments (red) and hexagonal ice Ih (purple) for (A) BOO-NN, (B) PointNet, (C) AE–GMM, and (D) this work. 
    `Density*' refers to probability densities that are normalized so that their maximum value is 1.0. (E) Mutual information for the 50 most informative descriptors, where green indicates selected descriptors and gray indicates unselected descriptors. Descriptors are selected for both their mutual information values and their correlation with previously chosen descriptors. 
    Schematics of descriptors and corresponding probability distributions from LDA and HDA environments for (F, G) $G_\text{H}^1$, (H, I) $q_3$, and (J, K) $\bar{q}_{11}$. 
    In the schematics, darker colors indicate higher descriptor values.}
    \label{fig:model_summary}
\end{figure*}

\begin{figure}[htb!]
    \centering
    \includegraphics[width=3in]{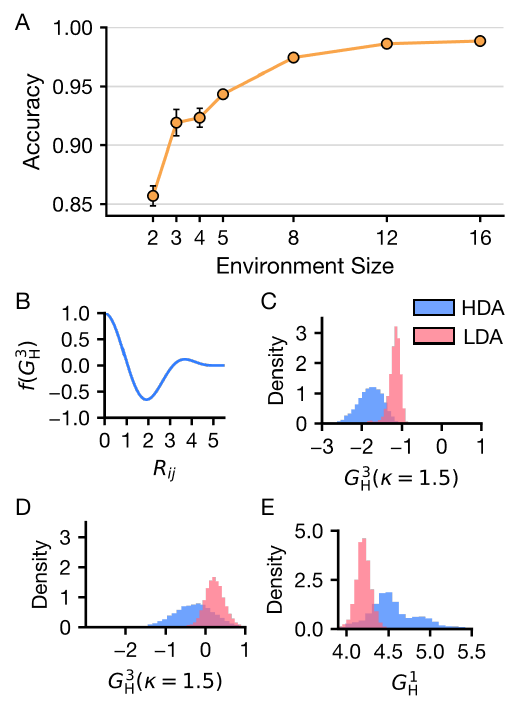}
    \captionsetup{justification=raggedright, singlelinecheck=false,font=small}
    \caption{
    \textbf{Amorphous state identity is encoded by length-scale dependent features.}
    (A) Classification accuracy as a function of environment size, where size is the number of neighboring water molecules. 
    Means and standard deviations are obtained from five-fold cross validation. 
    (B) Functional form of the per-neighbor contribution to $G_\text{H}^{3}$ as a function of interatomic distance.
    Distributions from HDA and LDA environments characterized by 
    (C) $G_\text{H}^{3}(\kappa=1.5)$ using environments of size 3, 
    (D) $G_\text{H}^{3}(\kappa=1.5)$ using environments of size 16, and 
    (E) $G_\text{H}^{1}$ using environments of size 3.}
    \label{fig:size_accuracy}
\end{figure}

We first assess how accurately and confidently our framework can classify LDA and HDA.
To evaluate performance, we benchmark against three approaches with distinct strategies: a neural network trained on bond-orientational order parameters (BOO-NN)~\cite{Martelli:2020, Faure:2024}, a coordinate-level architecture that learns representations directly from atomic positions (PointNet)~\cite{Defever:2019}, and an autoencoder combined with Gaussian mixture modeling (AE-GMM)~\cite{Boattini:2019}.
All models are trained using environments constructed from 16 nearest neighbors; full precision, recall, and accuracy metrics for DP\_MBpol and DP\_SCAN are reported in the SI (Tables S1 and S2).

All approaches achieve high classification accuracy, confirming that LDA and HDA environments are microscopically distinguishable.
Our method achieves 99.3\% \add{binary classification} accuracy, matching or surpassing all benchmarks and outperforming classifiers trained on BOO parameters or ACSFs alone, indicating that combining complementary descriptor types improves classification.
However, high classification accuracy alone does not guarantee reliability when applied to transformation trajectories, where configurations belonging to neither LDA nor HDA may arise.
To evaluate OOD detection, we examine how models classify ice Ih environments, which share LDA's tetrahedral coordination but possess long-range crystalline order.
BOO-NN and PointNet assign high LDA probabilities to many Ih configurations (Figures~\ref{fig:model_summary}A,B), reflecting overconfidence characteristic of neural-network classifiers that partition descriptor space without recognizing inputs outside the training distribution.
In contrast, AE-GMM and our method assign Ih probabilities orders of magnitude lower than genuine LDA structures (Figures~\ref{fig:model_summary}C,D), enabling reliable detection of novel configurations.
This capability is essential for analyzing transformation pathways where intermediate environments may exist.

\add{To test this more broadly, we evaluate OOD detection against 14 additional ice polymorphs spanning a wide range of pressures (SI Section S4). 
Using the complete set of weakly correlated descriptors, rather than only the five selected for LDA/HDA classification, substantially improves detection for most polymorphs, with several phases moving from near-zero to near-complete detection, though a subset of high-pressure phases remain only partially distinguishable from LDA/HDA even with the full descriptor set.
This residual structural overlap does not imply that these crystalline phases are present within LDA/HDA. 
Previous work has found no evidence of ice-like domains in computational LDA or HDA~\cite{Martelli:2018}, indicating that the classifier ambiguity for these phases instead reflects local structural similarity.}

\section*{Hierarchy of descriptor importance highlights key structural differences}

To identify which structural descriptors most effectively distinguish LDA and HDA, we examine the descriptors selected by mutual-information and correlation analysis in our probabilistic framework
(Figure \ref{fig:model_summary}E).
Notably, the ACSF descriptor $G^{1}_\text{H}$ exhibits the highest MI, demonstrating that radial density information is more discriminating than the orientational symmetry captured by BOO parameters, which have been the focus of prior work~\cite{Martelli:2020, Faure:2024}.
Many additional ACSF descriptors are nearly as informative as $G^{1}_\text{H}$ but are excluded due to their high correlation with it.
The remaining selected features are BOO parameters, which are consistent with those previously identified as changing significantly during LDA/HDA transitions~\cite{Faure:2024}.
These descriptors enable accurate classification for multiple water force fields (Tables S1 and S2), indicating that they capture structural differences intrinsic to the amorphous ices rather than force-field-specific artifacts.

To understand the physical significance of these descriptors, we examine how they encode structural differences between LDA and HDA.
The most informative descriptor, $G^{1}_{\text{H}}$, can be interpreted as a weighted hydrogen count in the local environment of an oxygen, up to $R_c = 5.0\,\text{\AA}$ or ca. 16 water molecules (Figure \ref{fig:model_summary}F). 
HDA environments exhibit substantially larger $G_{\text{H}}^1$ values than LDA (Figure \ref{fig:model_summary}G), consistent with HDA's higher bulk density.
Notably, local hydrogen density is more informative than the analogous oxygen-based descriptor $G_{\text{O}}^1$; this distinction emerges automatically from the descriptor selection.
The remaining selected descriptors consist of BOO parameters, which quantify spherical symmetries of varying order across different length scales (Figure \ref{fig:model_summary}H,J).
For all selected BOO descriptors, LDA environments exhibit larger values than HDA (Figure \ref{fig:model_summary}I,K).
This behavior is similar to that of the conventional tetrahedral order parameter $q$, for which average values are $q \approx 0.91$ for LDA and $q \approx 0.74$ for HDA~\cite{Szukalo:2025}. 
The HDA value is comparable to that of ambient liquid water, whose tetrahedrality is typically $q \approx 0.75$~\cite{Errington:2001}, whereas LDA lies much closer to the highly ordered tetrahedral structure of ice Ih, which exhibits $q \approx 0.98$~\cite{Chau:1998, Errington:2001}. 

Taken together, these descriptors reveal that local environments are distinguished primarily through variations in local density, with orientational symmetry playing a secondary but complementary role.
The prominence of density-based features is consistent with prior work identifying local density as correlated with effective order parameters for liquid water near its critical point~\cite{Foffi:2022}.
Within this low-dimensional descriptor space, individual molecular environments are classified as LDA or HDA based on their position along these axes, while environments falling outside the characteristic ranges of both phases are naturally identified as distinct.

\section*{LDA and HDA are resolvable within the first coordination shell due to interstitial hydrogen}
\label{sec:env_size}

A common feature of prior classification approaches for LDA and HDA is the use of structural information that extends beyond the first coordination shell, incorporating 16 nearest neighbors~\cite{Martelli:2020, Donkor:2024, Faure:2024}.
This perspective is reinforced by studies of supercooled liquid water, where clear separation between low- and high-density environments emerges only when structural information is averaged beyond the first shell~\cite{Donkor:2024}.
Motivated by these findings, we ask whether the same level of structural nonlocality is required to distinguish the glassy states.
To this end, we compute ACSFs and BOO parameters for local environments containing between 2 and 16 nearest neighbors. 
Here, we retain only local descriptors that do not incorporate information beyond the defined environment (\textit{i.e.}, excluding neighbor-averaged quantities such as $\bar{q}$) to directly quantify the minimal structural information required for classification.

Classification accuracy increases monotonically with environment size (Figure~\ref{fig:size_accuracy}A), at approximately 86\% with only two neighboring water molecules and plateauing above 97\% for eight or more neighbors.
These results indicate that the distinction between glassy LDA and HDA is encoded primarily in local packing motifs and not nanometer-scale domains.
That substantial phase-specific information is encoded within the immediate coordination shell is unexpected given prior work highlighting the necessity of second-shell information~\cite{Martelli:2020, Donkor:2024, Faure:2024}.
Rather, Figure~\ref{fig:size_accuracy}A shows that the first solvation shell contains much of the structural information needed to distinguish LDA and HDA, with larger environments extending toward the second solvation shell offering a modest further improvement in classification accuracy.

\add{These conclusions are necessarily bounded by the maximum environment size considered and do not exclude the possibility that structural differences between LDA and HDA  persist at longer range. 
The length-scale scan of Figure~\ref{fig:size_accuracy}A is computationally inexpensive and constitutes a useful diagnostic for any new application of this framework, revealing the environment size beyond which additional structural information ceases to improve discrimination; this saturation behavior may be system-specific. 
More broadly, the question addressed here is inherently local.
LDA and HDA are readily distinguishable by their bulk properties, and the non-trivial result is that this distinction is already encoded at the level of individual molecular environments.}

To understand what structural features most strongly delineate LDA and HDA at the smallest scales, we examine the most informative descriptors when considering only three neighboring water molecules.
We find that $G_\text{H}^{3}(\kappa = 1.5)$ is the most informative descriptor at this scale.
The functional form (Figure~\ref{fig:size_accuracy}B) takes negative values at distances between 1.5 and 3~\AA, corresponding to hydrogen-bonded neighbors.
HDA environments exhibit more negative values than LDA (Figure~\ref{fig:size_accuracy}C), indicating higher hydrogen density within hydrogen-bonding distance of the central molecule.
This is consistent with densification disrupting the tetrahedral hydrogen-bond network of LDA through insertion of interstitial water molecules into the first coordination shell.
The classifier effectively detects these interstitial hydrogens through $G_\text{H}^{3}$, enabling accurate LDA--HDA discrimination using only first-shell information.

However, the quality of the information provided by $G_\text{H}^{3}$ depends on spatial scale.
This is demonstrated by computing its value using 16 neighboring water molecules and observing the substantial overlap between distributions of LDA and HDA configurations (Figure~\ref{fig:size_accuracy}D).
\add{Thus, the ability of $G_\text{H}^{3}$ to discriminate LDA and HDA worsens with increasing the environment size.}
Conversely, $G_\text{H}^{1}$, the most informative descriptor for these larger environments (Figure \ref{fig:model_summary}F), provides little separation for environments of three neighbors (Figure~\ref{fig:size_accuracy}E).
These contrasting behaviors indicate that different structural correlations dominate at different length scales.
In particular, interstitial hydrogen insertion is most clearly distinctive at local scales, while the cumulative density differences captured by $G_\text{H}^{1}$ emerge only when averaging over larger environments.

\section*{Redistribution of local environments underlies transition and highlights first-order character}

We next consider the microscopic character of the LDA--HDA transformation.
Specifically, does it proceed through structurally distinct intermediate environments or via redistribution between locally LDA-like and HDA-like motifs?
Experimentally, LDA and HDA can be interconverted via isothermal compression and decompression, with signatures commonly interpreted as first-order-like phase transitions~\cite{Mishima:1984, Klotz:2005, Loerting:2006}. 
We apply our classifier to isothermal compression and decompression trajectories at $T = 80$~K for DP\_MBpol water~\cite{Szukalo:2025}, labeling each molecule as LDA, HDA, or outlier based on its local environment.

\begin{figure*}
    \centering
    \includegraphics[width=\textwidth]{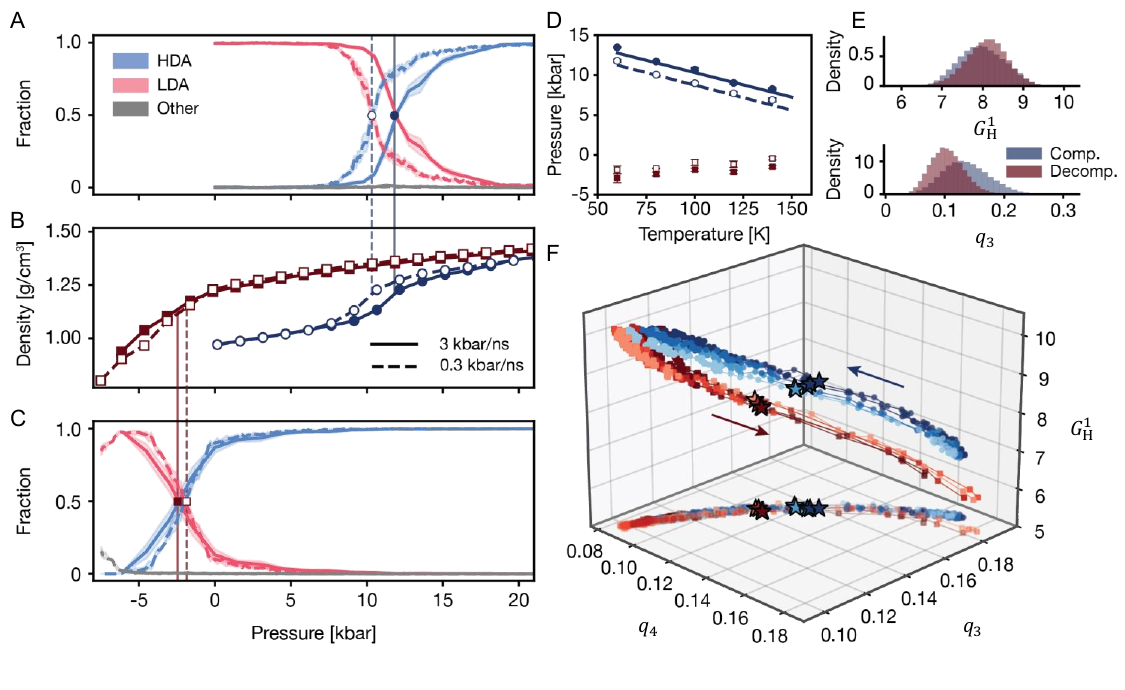}
    \captionsetup{justification=raggedright, singlelinecheck=false,font=small}
    \caption{
    \textbf{Compression and decompression of amorphous ices follow structurally distinct microscopic pathways without intermediate local environments.}
(\textbf{A}) Evolution of LDA (red), HDA (blue), and outlier (gray) populations during isothermal compression at $T = 80$~K for rates of 3~kbar/ns (solid) and 0.3~kbar/ns (dashed). 
Vertical lines indicate transition pressures, defined by equal LDA and HDA populations.
(\textbf{B}) Density--pressure equation of state extracted from compression (circles) and decompression (squares) trajectories for rates of 3~kbar/ns (filled) and 0.3~kbar/ns (open). 
(\textbf{C}) Analogous evolution of microscopic populations during decompression. 
(\textbf{D}) Out-of-equilibrium phase diagram deduced from local structural descriptors (markers with same meaning as in B) \add{based on compression/decompression runs at different temperatures.}  
Lines indicate LDA-to-HDA transition pressures from a thermodynamic criterion applied to the equation of state~\cite{Szukalo:2025}, which is reliable for compression but not decompression.
(\textbf{E}) Distributions of $G_{\mathrm{H}}^{1}$ (top) and $q_3$ (bottom) for environments sampled during compression (blue) and decompression (red) within a common density window ($1.15$--$1.25$~g/cm$^3$).
(\textbf{F}) \add{
Mean local-environment descriptors ($q_3$, $q_4$, $G^1_\mathrm{H}$) along compression (blue) and decompression (red) pathways for separate isothermal MD simulations at temperatures from 60 to 140 K. 
The upper set of curves shows trajectories in the full $q_3$--$q_4$--$G^1_\mathrm{H}$ space; the lower set shows the same trajectories projected onto the $q_3$--$q_4$ plane.
Stars mark transition points; darker shades correspond to lower temperatures. 
See SI Figures~S2 and S3 for 2D projections.}}
    \label{fig:transitions}
\end{figure*}

Figure~\ref{fig:transitions}A--C summarizes the transformation behavior.
The LDA$\rightarrow$HDA transition unfolds continuously over approximately 8--16~kbar during compression.
The reverse transformation requires negative pressures, with LDA and HDA populations coexisting over a broad pressure range before LDA is fully recovered, which is indicative of pronounced hysteresis.
Crucially, the transformation proceeds entirely through changes in the relative populations of LDA- and HDA-like environments, with no intermediate motifs observed.
This population-based picture is consistent with LDA and HDA corresponding to separate regions on the underlying potential energy landscape~\cite{Giovambattista:2003, Giovambattista:2016} and provides direct molecular-scale evidence for the first-order-like phase transition character of the LDA-HDA transformation.

\add{The population-redistribution picture also connects to the phenomenology of fluid polyamorphism developed by Anisimov and coworkers \cite{Anisimov:2018}, which distinguishes between a ``discrete'' mechanism, driven by interconversion between two structurally distinct molecular states, and a ``continuous'' mechanism arising from nonideality in the Gibbs energy, both of which can produce identical macroscopic signatures. 
The molecule-resolved probabilities $P_{\mathrm{LDA},i}$ and $P_{\mathrm{HDA},i}$ computed here are analogous microscopic realizations in the glass state, of the interconversion fractions of the liquid state, central to that framework.
The absence of intermediate environments along the LDA-HDA transformation pathway indicates that the discrete mechanism operates in water's amorphous ices.
Our results are consistent with theoretical descriptions of water based on two-state models~\cite{Tanaka:2000, Tanaka:2020}}

\section*{Local structural motifs reliably map LDA/HDA boundaries}
\label{sec:kinetic}

We next use molecule-resolved classification to identify LDA/HDA transition pressures across a range of temperatures, defining the transition as the point of equal LDA and HDA populations.
Conventionally, the LDA--HDA transition pressures are estimated from the rate-dependent density--pressure equation of state (EoS), where the LDA$\rightarrow$HDA transformation appears as a maximum in the rate of density change with respect to pressure (an inflection point in the density-pressure EoS).
However, the reverse HDA$\rightarrow$LDA transition lacks clear thermodynamic signatures in the EoS, precluding  identification of its transition pressure~\cite{Szukalo:2025}. 

For compression trajectories, classifier-derived and EoS-based transition pressures agree quantitatively (Figure~\ref{fig:transitions}D, circles and lines), reflecting strong coupling between local structural reorganization and macroscopic densification.
For decompression, the classifier robustly identifies HDA$\rightarrow$LDA transition pressures even where the EoS varies continuously without inflection (Figure~\ref{fig:transitions}D, squares), enabling construction of complete out-of-equilibrium phase diagrams, including the decompression spinodal previously inaccessible from thermodynamic analysis~\cite{Szukalo:2025}.
Furthermore, substantial hysteresis is observed, with decompression transitions occurring at negative pressures and differing from compression transition pressures by 15--25~kbar, consistent with large kinetic barriers characteristic of first-order-like transformations.

\section*{Structural pathways of pressure-induced transformations exhibit hysteresis}

Given the substantial hysteresis in transition pressures, we next ask whether compression and decompression follow identical structural pathways by tracking mean values of $G_{\mathrm{H}}^{1}$, $q_3$, and $q_4$ along trajectories across temperatures from 60 to 140~K.
The distributions of $G_{\mathrm{H}}^{1}$ and $q_3$ sampled within a common density window of $1.15$--$1.25$~g/cm$^3$ (Figure~\ref{fig:transitions}E) directly demonstrate structural hysteresis.
For $q_3$, compression and decompression distributions are clearly separated, demonstrating that the forward and reverse transformations traverse structurally distinct regions of \add{local} configuration space.
More subtly, $G_{\mathrm{H}}^{1}$ also differs between pathways despite identical bulk densities, a nontrivial result given that $G_{\mathrm{H}}^{1}$ reflects local interstitial hydrogen density and might naively be expected to track global density alone.

In the $q_3$--$G_{\mathrm{H}}^{1}$ plane (Figure~\ref{fig:transitions}F), compression and decompression follow clearly distinct paths across all temperatures.
Compression proceeds in two stages: \emph{(i)} initial densification with little loss of orientational order (increasing $G_{\mathrm{H}}^{1}$ at nearly constant $q_3$), reflecting elastic compression of the hydrogen-bond network, followed by \emph{(ii)} collapse of tetrahedral order as densification continues into the HDA basin.
Temperature primarily affects the second stage, with higher temperatures allowing tetrahedral order to collapse at slightly higher hydrogen coordination.
Decompression follows a qualitatively different sequence: density decreases substantially before any recovery of tetrahedral order, with $q_3$ increasing only after the system has expanded well into the low-density regime.

Strikingly, when $G_{\mathrm{H}}^{1}$ is replaced by $q_4$, compression and decompression trajectories collapse onto nearly identical curves, and the transformation appears reversible.
This pattern holds for any combination of BOO parameters alone (SI Figures~S2 and S3), demonstrating that the microscopic interpretation of amorphous transformations depends critically on descriptor choice; local hydrogen density reveals pronounced hysteresis that orientational order parameters alone entirely obscure.

\section*{Different force fields produce locally distinct LDA phases}
\label{sec:ff}

\begin{table}
\captionsetup{justification=raggedright, singlelinecheck=false,font=small}
\caption{
Most informative local structural descriptors for resolving DP\_SCAN and DP\_MBpol local environments.}
\centering
\begin{tabular}{cc}
\toprule
Feature & Mutual Information \\
\midrule
$G^{2}_{\text{O}}(\eta=2.5,R_{s}=1.0)$ & 0.220\\
$G^{2}_{\text{H}}(\eta=1.0,R_{s}=1.0)$ & 0.188\\
$G^{5}_{\text{H,H}}(\eta=4.5,\zeta=0.5,\lambda=-1.0)$ & 0.142\\
$G^3_\text{H}(\kappa=0.5)$ & 0.141\\
$G^3_\text{O}(\kappa=2.0)$ & 0.117\\
\bottomrule
\end{tabular}
\label{table:feat_scan_mbpol}
\end{table}

\begin{figure}
    \centering
    \includegraphics[width=3.5in]{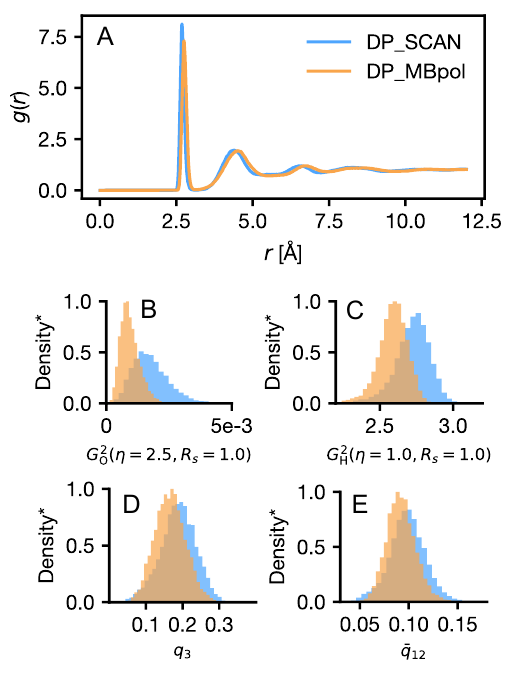}
    \captionsetup{justification=raggedright, singlelinecheck=false,font=small}
    \caption{
    \textbf{Radial packing variations highlight differences between quantum-chemistry-derived water models.} 
    (A) Oxygen–oxygen radial distribution functions for LDA at $T = 80$~K and $P = 1$~bar.  
    (B--E) Distributions of select ACSF and BOO descriptors for LDA generated by each model across temperatures from 60 to 140~K at $P = 1$~bar. 
    All probability densities are normalized such that the maximum value of either distribution is 1.0, with both distributions scaled by the same factor (denoted as Density$^*$).
    The blue and orange colors represent the DP\_SCAN- and DP\_MBpol models, respectively.}
    \label{fig:scan_mbpol}
\end{figure}

As a further test, we now ask whether the same titular amorphous phase, generated by different force fields, is structurally distinguishable.
A classifier trained to distinguish DP\_SCAN and DP\_MBpol LDA configurations achieves $77.5 \pm 0.3\%$ accuracy on unseen environments, confirming that the two models produce measurably different local structures.
Structural differences between HDA configurations are considered in the SI, Fig~S4.
The most discriminating features are exclusively ACSFs, with no BOO parameters within the top five (Table~\ref{table:feat_scan_mbpol}), indicating that the models differ primarily in radial packing rather than orientational symmetry.
This is consistent with oxygen--oxygen radial distribution functions at $T = 80$~K, which show a slightly higher short-range peak for DP\_SCAN (Figure~\ref{fig:scan_mbpol}A), and with ACSF descriptors that confirm systematically higher local densities for both oxygen and hydrogen in DP\_SCAN environments (Figures~\ref{fig:scan_mbpol}B,C), reflecting the tendency of SCAN-based potentials to overstabilize close-contact configurations~\cite{Zhang:2021pd, Szukalo:2025}.

Orientational descriptors reveal more subtle differences.
DP\_MBpol produces slightly lower tetrahedral order and greater variability in local orientational environments (Figures~\ref{fig:scan_mbpol}D,E), indicating a more distorted hydrogen-bond network despite both models capturing the overall tetrahedral topology of LDA.
These results demonstrate that environment-resolved structural analysis provides a sensitive lens for comparing machine-learned potentials as such models proliferate.

\section*{Implications}

The findings reported here advance the structural picture of water's amorphous ices in several respects.
The locality of the LDA--HDA distinction, encoded within the first coordination shell, challenges the prevailing view that second-shell correlations are necessary to resolve amorphous water environments~\cite{Martelli:2020, Donkor:2024}.
In vitrified glasses, density fluctuations are arrested in distinct configurational basins, and differences in interstitial occupancy become sufficiently pronounced that first-shell descriptors alone suffice, suggesting that the requisite length scale for structural classification is state-dependent.
\add{This state-dependence suggests that extending the framework to equilibrium liquid environments where broader thermal fluctuations are expected to increase configuration overlap may require a larger or different descriptor set. 
Classification of LDL and HDL environments as the liquid--liquid 
critical point is approached represents a compelling direction for future work.}
The absence of intermediate structural populations along transformation pathways, established here with a classifier capable of explicitly detecting OOD configurations, constrains the energy landscape picture of glass--glass interconversion~\cite{Giovambattista:2016} and provides molecular-scale evidence for first-order character even where macroscopic observables vary continuously.
\add{Additionally, the molecule-resolved classification demonstrated here offers a general microscopic strategy for distinguishing the discrete and continuous mechanisms of polyamorphism~\cite{Anisimov:2018}.}

More broadly, the finding that orientational order parameters alone mask the hysteresis revealed by local hydrogen density carries a fundamental cautionary implication.
The microscopic interpretation of amorphous transformations depends critically on descriptor choice, and analyses based on a single class of order parameter risk overlooking essential structural physics.
\add{This descriptor-dependence extends to OOD detection itself.
Broadening the descriptor set used to flag anomalous environments substantially improved sensitivity to several high-pressure ice polymorphs.}
The probabilistic framework introduced here is applicable to any system in which distinct amorphous or disordered phases lack obvious microscopic distinguishing features.
The environment-resolved comparison of machine-learned potentials illustrates one such application, revealing force-field-specific structural signatures invisible to conventional metrics.
As data-driven interatomic potentials become increasingly prevalent, such analyses will be essential for connecting microscopic structure to emergent condensed-phase behavior.

\section*{Methods}

We analyze MD trajectories of LDA and HDA ice generated using two machine-learned interatomic potentials: DP\_SCAN, trained on DFT data with the SCAN functional~\cite{Sun:2015, Zhang:2021pd}, and DP\_MBpol, trained on the many-body polarizable MB-pol potential parameterized against coupled-cluster calculations~\cite{Babin:2013, Babin:2014, Medders:2014, Bore:2023}.
All simulations were performed using LAMMPS (2~Aug~2023)~\cite{Thompson:2022} interfaced with DeePMD-kit (v3.0.0)~\cite{Zhang:2018, Wang:2018}; full simulation protocols are provided in Ref.~\citenum{Szukalo:2025}.

The training and evaluation dataset comprises 10,000 molecular environments, equally divided between LDA and HDA.
Configurations were deliberately selected from thermodynamic conditions where phase identity is unambiguous, avoiding the transition region.
LDA environments were drawn from isobaric quenches at temperatures at least 30~K below the glass transition and from isothermal compression of LDA at pressures at least 3~kbar below its transformation pressure to HDA.
HDA environments were taken from compression trajectories at least 3~kbar above the LDA-to-HDA transformation pressure, as well as from decompression trajectories restricted to densities exceeding that of recovered HDA at 80~K and 1~bar, to exclude partially transformed structures.
An additional 1,000 hexagonal ice~(Ih) configurations, which share the tetrahedral hydrogen-bond network of LDA but possess long-range crystalline order, were curated as a benchmark for out-of-distribution~(OOD) detection.
All classification results are validated using five-fold cross-validation.

Each molecule's local environment is characterized using two complementary, rotationally invariant descriptor families.
Atom-centered symmetry functions~(ACSFs)~\cite{Behler:2011} encode radial and angular correlations of interatomic distances through a set of radial ($G^1$, $G^2$, $G^3$) and angular ($G^4$, $G^5$) functions, computed separately for oxygen and hydrogen neighbors to retain sensitivity to hydrogen-bond motifs and proton arrangements not visible from oxygen positions alone.
Bond-orientational order~(BOO) parameters~\cite{Steinhardt:1983}, including locally averaged variants~\cite{Lechner:2008}, quantify the orientational symmetry of the surrounding molecular arrangement via spherical harmonics.
Unless otherwise stated, local environments are defined by the 16 nearest neighboring water molecules, with a cutoff radius of $R_\text{c} \approx 5.0$~\AA, consistent with prior work~\cite{Martelli:2020, Donkor:2024, Faure:2024}.
Full descriptor definitions and parameter sets are provided in the SI.

The most discriminative descriptors are identified through a two-stage selection procedure.
First, each descriptor is ranked by its mutual information~(MI) with the class label~\cite{Shannon:1948}, retaining those with MI values at least 10\% of the maximum observed value.
Second, a greedy correlation filter removes redundant descriptors.
Proceeding in order of decreasing MI, a descriptor is retained only if its Pearson correlation with all previously retained descriptors satisfies $|r| < 0.8$.
This yields a compact set of informative, approximately uncorrelated descriptors without requiring prior assumptions about which structural motifs are important.
\add{For classification, we use the five most informative of these descriptors (Fig.~\ref{fig:model_summary}E); for OOD detection (below), we instead use the complete correlation-filtered descriptor set, as additional weakly correlated descriptors provide independent axes along which an anomalous environment may be identified and thereby maximize outlier detection sensitivity.}

\add{Classification proceeds by modeling the class-conditional distribution of each selected descriptor independently using Gaussian kernel density estimation~(KDE), with bandwidths optimized by five-fold cross-validation on the provided training data.}
For a new environment, the joint likelihood for each class is computed as the product of per-descriptor likelihoods under a na\"{i}ve Bayes assumption, and the environment is assigned to the highest-likelihood class.
A key advantage of this formulation is explicit OOD detection.
\add{If an environment falls outside of the LDA and HDA reference distributions along any of the descriptors retained by the correlation filter (defined by a
kernel density estimate below $5 \times 10^{-4}$), then it is flagged as
structurally novel rather than forced into a known category.
We note that OOD detection in this framework is inherently relative to the structural descriptors provided.
Structures that are anomalous along alternative descriptor axes may not be reliably flagged.}
Five-fold cross-validation is used to obtain robust estimates of classification accuracy. 
Full mathematical details of the probabilistic model are provided in the SI.

\section*{Code Availability}
All code required to recreate the results above can be obtained at \texttt{https://github.com/webbtheosim/amorphous-ice}.

\begin{acknowledgments}
Q.M.G. acknowledges support from the National Science Foundation Graduate Research Fellowship Program under Grant No. DGE-2039656. 
R.J.S., P.G.D., and M.A.W acknowledge support from the ``Chemistry in Solution and at Interfaces'' (CSI) Center funded by the U.S. Department of Energy through Award No. DE-SC0019394. 
N.G. is thankful to the NSF-CREST Center for Interface Design and Engineered Assembly of Low Dimensional systems (IDEALS; grant numbers HRD-1547380 and HRD-2112550).
Simulations and analyses were performed using resources from Princeton Research Computing at Princeton University, which is a consortium led by the Princeton Institute for Computational Science and Engineering (PICSciE) and Office of Information Technology's Research Computing.\\
\end{acknowledgments}

\bibliography{bibliography}

\clearpage
\appendix
\onecolumngrid

\begin{center}
{\Large\textbf{Supporting Information for ``A Local Structural Basis to Resolve Amorphous Ices"}}\\[1em]
 
Quinn M. Gallagher$^{1,*}$,
Ryan J. Szukalo$^{2,*,\dagger}$,
Nicolas Giovambattista$^{3,4}$,
Pablo G. Debenedetti$^{1}$,
Michael A. Webb$^{1,\dagger}$\\
[0.5em]
 
{\small
$^{1}$Department of Chemical and Biological Engineering, Princeton University, Princeton, NJ 08540, United States\\
$^{2}$Department of Chemistry, Princeton University, Princeton, NJ 08540, United States\\
$^{3}$Department of Physics, Brooklyn College of the City University of New York, Brooklyn, New York 11210, United States\\
$^{4}$Ph.D. Programs in Physics and Chemistry, The Graduate Center of the City University of New York, New York, New York 10016, United States\\
 
\vspace{1em}
$^{*}$These authors contributed equally to this work.\\
 
\vspace{1em}
$^{\dagger}$Corresponding Authors: mawebb@princeton.edu, rszukalo@princeton.edu
}
\end{center}

\setcounter{section}{0}
\setcounter{figure}{0}
\setcounter{table}{0}
\setcounter{page}{1}

\renewcommand{\thesection}{S\arabic{section}}  
\renewcommand{\thetable}{S\arabic{table}}  
\renewcommand{\thefigure}{S\arabic{figure}}

\noindent\textbf{Contents}
\vspace{0.5em}

\noindent\hyperref[sec:S1]{S1. Methods} \dotfill \pageref{sec:S1}

\noindent\hyperref[sec:S2]{S2. Benchmarking Model Performance for DP\_SCAN Environments} \dotfill \pageref{sec:S2}

\noindent\hyperref[sec:S3]{S3. Out-of-distribution Predictions for DP\_SCAN and DP\_MBpol Environments} \dotfill \pageref{sec:S3}

\noindent\hyperref[sec:S4]{S4. Out-of-distribution Prediction for Alternative Phases of Ice} \dotfill \pageref{sec:S4}

\noindent\hyperref[sec:S5]{S5. Sensitivity of Structural Pathways to Selected Feature Spaces} \dotfill \pageref{sec:S5}

\noindent\hyperref[sec:S6]{S6. Differences between DP\_SCAN- and DP\_MBpol-generated HDA Environments} \dotfill \pageref{sec:S6}

\newpage

\section{Methods}
\label{sec:S1}

Figure 1 in the main text illustrates our framework that combines interpretable structural descriptors with probabilistic classification to identify local collective variables that distinguish amorphous phases.
First, each molecule's local environment is represented using atom-centered symmetry functions (ACSFs) and bond-orientational order (BOO) parameters, which together provide a comprehensive, symmetry-invariant characterization of local structure (Step 1).
Second, a two-stage feature selection procedure uses mutual information (MI) to rank descriptors by discriminative power and removes redundant descriptors through correlation analysis (Step 2).
Third, class-conditional probability distributions of the selected descriptors are then modeled, enabling transparent probabilistic classification (Step 3).
Finally, the joint probability for each class is used for molecule-by-molecule state assignment and explicit detection of out-of-distribution configurations, i.e., \add{molecular environments} that cannot be classified as LDA-like or HDA-like (Step 4).
This approach automatically identifies which structural descriptors best distinguish the amorphous ices without requiring \textit{a priori} knowledge of the relevant order parameters, providing both high classification accuracy and physical interpretability.
The following sections detail each step of this procedure.

\subsection{Structural Descriptors}
\label{sec:descriptors}

We consider two complementary, interpretable descriptor families that together provide a comprehensive characterization of microscopic environments.
ACSFs encode the local environment through radially and angularly resolved correlations of interatomic distances~\cite{Behler:2011}.
BOO parameters~\cite{Steinhardt:1983} provide a complementary description that quantifies the orientational symmetry of the surrounding molecular arrangement and have been widely used to characterize amorphous and crystalline ices~\cite{Martelli:2020, Faure:2024}.  
Together, these descriptors form a symmetry-invariant and physically grounded representation of each molecule's local environment (Fig. 1, Step 1).

\subsubsection{Atom-Centered Symmetry Functions}

ACSFs are computed using several distinct functional forms that capture complementary aspects of local atomic environments through radially and angularly resolved correlations.
All ACSF functions are centered on oxygen atoms (central oxygen atom $i$), with contributions included from both oxygen and hydrogen neighbors to ensure sensitivity to variations in hydrogen-bond motifs and proton arrangements not visible from oxygen positions alone.
Throughout, we denote $\mathbf{r}_{ij}$ as the displacement vector from central atom $i$ to neighboring atom $j$ and $R_{ij}=\|\mathbf{r}_{ij}\|$ as the corresponding scalar distance.
All ACSF functions employ a smooth cutoff function $f_\text{c}(R_{ij})$:

\begin{equation}
f_\text{c}(R_{ij}) = \begin{cases}
0.5 \left[\cos\left(\frac{\pi R_{ij}}{R_\text{c}}\right) + 1\right] & \text{if } R_{ij} \leq R_\text{c} \\
0 & \text{if } R_{ij} > R_\text{c},
\end{cases}
\end{equation}

\noindent This cutoff function smoothly decays from 1 at $R_{ij} = 0$ to 0 at $R_{ij} = R_\text{c}$, with both the function and its first derivative vanishing continuously at the cutoff radius.
We define the local environment around a central oxygen $i$ by the 16 nearest neighboring water molecules.
Let $\mathcal{N}_{\mathrm{mol}}(i)$ denote this set of 16 neighbor molecules.
We then define the corresponding sets of neighboring atoms as $\mathcal{N}_{\mathrm{O}}(i)$ and $\mathcal{N}_{\mathrm{H}}(i)$, the set of the 16 neighboring oxygen atoms and 32 neighboring hydrogen atoms belonging to the molecules in $\mathcal{N}_{\mathrm{mol}}(i)$.
Unless otherwise stated, we set $R_\text{c} \approx 5.0$\,\AA\ and always ensure that all atoms in these neighbor sets satisfy $R_{ij}\le R_\text{c}$ (\textit{i.e.}, the neighbor list is defined by neighbor count).

Three types of radial symmetry functions provide complementary representations of the local density. 
The simplest radial function, $G^1$, provides a weighted count of neighboring atoms:

\begin{equation}
G^1 = \sum_{j \in \mathcal{N}_\gamma(i)} f_\text{c}(R_{ij}),
\label{eq:g1}
\end{equation}

\noindent where $\gamma\in\{\mathrm{O},\mathrm{H}\}$ specifies the neighbor species set.
In particular, we compute separate $G^1$ functions for oxygen and hydrogen neighbors, denoted $G^1_\text{O}$ and $G^1_\text{H}$.

The next function, $G^2$, focuses on the local density at a particular distance from the central atom,

\begin{equation}
G^2(\eta, R_\text{s}) = \sum_{j \in \mathcal{N}_\gamma(i)} \exp\left[-\eta(R_{ij} - R_\text{s})^2\right] f_\text{c}(R_{ij}),
\end{equation}

\noindent where $\eta$ controls the width of the Gaussian and $R_\text{s}$ shifts its center.
By using multiple $G^2$ functions with different $\eta$ and $R_\text{s}$ values, these functions can construct a radially resolved representation of the local density.

The third function, $G^3$, utilizes cosine damping:

\begin{equation}
G^3(\kappa) = \sum_{j \in \mathcal{N}_\gamma(i)} \cos(\kappa R_{ij}) f_\text{c}(R_{ij}),
\end{equation}

\noindent where $\kappa$ adjusts the period length. 
Using multiple $G^3$ functions with different $\kappa$ values provides a Fourier-like decomposition of the radial distribution.

Two types of angular functions capture information about three-body configurations and local orientational order.
This is achieved by considering angles centered at atom $i$ formed by neighbor atoms $j$ and $k$,

\begin{equation}
\theta_{jik} = \arccos\!\left(\frac{\mathbf{r}_{ij} \cdot \mathbf{r}_{ik}}{R_{ij} R_{ik}}\right).
\end{equation}
The two functions take the form (summing over unordered neighbor pairs $j<k$):
\begin{equation}
G^4(\eta, \zeta, \lambda) = 2^{1-\zeta} \sum_{j \in \mathcal{N}_\alpha(i)} \sum_{\substack{k \in \mathcal{N}_\beta(i) \\ k>j}}
(1 + \lambda \cos\theta_{jik})^\zeta 
\times \exp[-\eta(R_{ij}^2 + R_{ik}^2 + R_{jk}^2)] f_\text{c}(R_{ij}) f_\text{c}(R_{ik}) f_\text{c}(R_{jk}),
\end{equation}
and
\begin{equation}
G^5(\eta, \zeta, \lambda) = 2^{1-\zeta} \sum_{j \in \mathcal{N}_\alpha(i)} \sum_{\substack{k \in \mathcal{N}_\beta(i) \\ k>j}}
(1 + \lambda \cos\theta_{jik})^\zeta
\times \exp[-\eta(R_{ij}^2 + R_{ik}^2)] f_\text{c}(R_{ij}) f_\text{c}(R_{ik}).
\end{equation}
\noindent Here, $\alpha,\beta \in \{\mathrm{O},\mathrm{H}\}$ specify the neighbor species sets used for $j$ and $k$.
The parameter $\zeta$ controls the angular resolution, with larger values emphasizing specific angular arrangements, while $\lambda \in \{-1, 1\}$ shifts the sensitivity to linear ($\lambda = -1$) versus bent ($\lambda = 1$) configurations.
The parameter $\eta$ controls the radial extent over which angular correlations are evaluated.
The key distinction between $G^4$ and $G^5$ lies in their radial dependencies.
$G^4$ includes the neighbor--neighbor distance $R_{jk}$, restricting contributions to compact triplets where all three interatomic distances are small, while $G^5$ omits this constraint and thus captures angular correlations over a wider range of geometries.
These angular functions are computed over all possible species combinations of the three atoms involved, with parameter sets chosen to capture angular correlations at different radial extents.

\subsubsection{Bond-Orientational Order Parameters}

BOO parameters provide a complementary description of local structure by quantifying orientational symmetry through spherical harmonics \cite{Steinhardt:1983}.
Unlike ACSFs which are explicitly constructed as sums over neighbors, BOO parameters characterize the angular distribution of the local environment in a rotationally invariant manner.
For each oxygen atom $i$, we use the same oxygen neighbor list as above and define $N_\text{b}=|\mathcal{N}_{\mathrm{O}}(i)|=16$ neighboring oxygen atoms.
The complex-valued order parameter $q_{lm}$ is computed as:

\begin{equation}
q_{lm}(i) = \frac{1}{N_\text{b}} \sum_{j \in \mathcal{N}_{\mathrm{O}}(i)} Y_{lm}(\theta_{ij}, \phi_{ij}),
\end{equation}

\noindent where $Y_{lm}$ are spherical harmonics of degree $l$ and order $m$ and $(\theta_{ij}, \phi_{ij})$ are the polar and azimuthal angles of the vector $\mathbf{r}_{ij}$ connecting oxygen atoms $i$ and $j$.
The Steinhardt BOO parameter $q_l$ is obtained by combining all $m$ components for a given $l$:

\begin{equation}
q_l(i) = \sqrt{\frac{4\pi}{2l+1} \sum_{m=-l}^{l} |q_{lm}(i)|^2}.
\label{eq:stein}
\end{equation}

\noindent Low values of $l$ (e.g., $l = 3, 4$) are sensitive to tetrahedral and other low-symmetry arrangements, while higher $l$ values capture more complex orientational motifs.

We also include third-order descriptors $w_l$, which provide additional sensitivity to the shape of the orientational distribution:

\begin{equation}
w_l(i) = \frac{\sum_{m_1+m_2+m_3=0} C_{l}^{m_1 m_2 m_3} \, q_{lm_1}(i)\, q_{lm_2}(i)\, q_{lm_3}(i)}{\left(\sum_{m=-l}^{l} |q_{lm}(i)|^2\right)^{3/2}},
\label{eq:wl}
\end{equation}

\noindent where $C_{l}^{m_1 m_2 m_3}$ are the Wigner 3-$j$ coupling coefficients and the sum runs over all triplets $(m_1, m_2, m_3)$ satisfying $m_1 + m_2 + m_3 = 0$.
The $w_l$ parameters are particularly useful for distinguishing structures with similar $q_l$ values but different higher-order orientational correlations.

Additionally, we compute locally averaged variants that incorporate information from the neighborhood surrounding each molecule~\cite{Lechner:2008}.
The locally averaged complex order parameter $\bar{q}_{lm}(i)$ is defined as:

\begin{equation}
\bar{q}_{lm}(i) = \frac{1}{N_\text{b} + 1} \left[ q_{lm}(i) + \sum_{j \in \mathcal{N}_{\mathrm{O}}(i)} q_{lm}(j) \right].
\label{eq:q_bar_lm}
\end{equation}

\noindent This averaging procedure includes contributions from both the central molecule $i$ and its $N_\text{b}$ oxygen neighbors.
From $\bar{q}_{lm}(i)$, we compute the rotationally invariant 

\begin{equation}
\bar{q}_l(i) = \sqrt{\frac{4\pi}{2l+1} \sum_{m=-l}^{l} |\bar{q}_{lm}(i)|^2}.
\label{eq:q_bar}
\end{equation}

\noindent The locally averaged descriptors $\bar{q}_l$ provide enhanced discrimination between crystal structures compared to their unaveraged counterparts $q_l$, as the averaging reduces thermal fluctuations while preserving sensitivity to orientational order~\cite{Lechner:2008}.
The corresponding locally averaged third-order descriptors $\bar{w}_l$ are computed using Eq.~\eqref{eq:wl} but with all $q_{lm}(i)$ substituted by $\bar{q}_{lm}(i)$.

Together with the ACSF descriptors, this comprehensive set of structural descriptors provides a detailed, physically interpretable representation of each oxygen-centered local environment.

\subsection{Automatic Feature Selection}

We extract the most informative descriptors from the full set using an automated ranking procedure (Fig.~1, Step 2).
We define local environments using the 16 nearest neighbors, consistent with prior work~\cite{Martelli:2020, Donkor:2024, Faure:2024}.
We assess the ability of each descriptor to discriminate between classes (LDA-like and HDA-like) using mutual information (MI)~\cite{Shannon:1948}. 
MI quantifies the reduction in uncertainty about the class label upon observing a descriptor value and thus provides a model-free measure of discriminative power.
For a feature $X$ and class label $Y$, where $x$ and $y$ denote particular values of $X$ and $Y$ respectively, the MI is defined as:

\begin{equation}
\text{MI}(X; Y) = \sum_{x \in X} \sum_{y \in Y} p(x, y) \log \frac{p(x, y)}{p(x)p(y)},
\end{equation}

\noindent where $p(x, y)$ is the joint probability density of feature $X$ and class label $Y$, and $p(x)$ and $p(y)$ are the corresponding distributions.
We rank features by MI and retain those with MI values of at least 10\% of the maximum observed MI, such that all features are expected to possess useful information.

To ensure that selected descriptors contribute distinct structural information,
we use a greedy high-correlation filter. 
Proceeding in order of decreasing MI, each descriptor is retained only if its Pearson correlation ($r$) with all previously retained descriptors satisfies $|r| < 0.8$.
The outcome is a set of informative and roughly uncorrelated descriptors, \add{which we denote $\mathcal{D} = \{1,\dots,M\}$}.
Because the procedure operates on arbitrary descriptor inputs, it enables data-driven identification of discriminative descriptors without requiring prior assumptions about which structural motifs are important.
\add{We retain $\mathcal{D}$ in its entirety for out-of-distribution detection (see Probabilistic Model, below), since each additional weakly correlated descriptor offers an independent opportunity to identify an anomalous environment and thus maximizes detection sensitivity. 
For classification, however,}
the number of descriptors retained \add{from $\mathcal{D}$} is a user-specified parameter; here we select \add{the $m=5$ most informative descriptors, denoted $\mathcal{D}_\text{cls} = \{1,\dots,m\} \subset
\mathcal{D}$,} balancing classification accuracy with physical interpretability. 
The full set of ACSF and BOO descriptors considered as candidates spans all functional forms defined in Section~\ref{sec:descriptors}, and the specific descriptors selected for LDA/HDA classification are shown in Figure~2E of the main text.

\subsection{Probabilistic Model}

The second component of our framework is a probabilistic classifier that assigns a label to each atomic environment as a calibrated likelihood of belonging to each structural class (Fig.~1, Step 3). 
Rather than learning a decision boundary, we model the probability distribution of each selected descriptor for each class.

To avoid index ambiguity, we use $a\in\{1,\dots,m\}$ to index the selected
$m$ features \add{ in $\mathcal{D}_\text{cls}$ used for classification} and
$n\in\{1,\dots,N_y\}$ to index training samples within class $y$.
For each feature $a$ and class $y$, we estimate the corresponding one-dimensional probability density $p_{a,y}(x)$ of observing value $x$ using Gaussian kernel density estimation (KDE).
In $k$-fold cross-validation, the dataset is partitioned into $k$ equal subsets.
The model is trained on $k-1$ subsets and evaluated on the held--out subset, with this procedure repeated $k$ times so that every sample serves as a validation point exactly once. 
Performance metrics are then averaged across all $k$ folds, providing a robust estimate of generalization performance with reduced sensitivity to any particular data split. 
Throughout this work, we use $k = 5$ (five-fold cross-validation).
Kernel bandwidths $\sigma_{a}$ are selected via a grid search that maximizes the marginal log-likelihood estimated by five-fold cross-validation.

Each KDE is normalized such that $\int_{-\infty}^{\infty} p_{a,y}(x)\,\mathrm{d}x = 1$, where

\begin{equation}
p_{a,y}(x) = \frac{1}{N_y} \sum_{n = 1}^{N_y} \frac{1}{\sqrt{2\pi\sigma_a^2}} \exp \left[ \frac{-(x - x_{a,n})^2}{2\sigma_a^2} \right].
\end{equation}

\noindent Here, $N_y$ is the number of samples for class $y$ and $x_{a,n}$ are the observed values of feature $a$ within that class.

Given an observed feature value $x_a^*$, we define the probability within a kernel window of half-width $0.5\sigma_a$ centered at $x_a^*$ as

\begin{equation}
\pi_{a,y}(x_a^*) = \int_{x_a^* - 0.5\sigma_a}^{x_a^* + 0.5\sigma_a} p_{a,y}(x)\,\mathrm{d}x.
\end{equation}

For an environment characterized by $m$ selected features, $\vec{x}^* = (x_1^*, x_2^*, \dots, x_m^*)$, we compute a class-conditional likelihood score under a na\"ive Bayes assumption (Fig.~1, Step 4):

\begin{equation}
\mathcal{P}_y(\vec{x}^*) = \prod_{a=1}^{m} \pi_{a,y}(x_a^*).
\end{equation}

\noindent Each environment is assigned the class with the largest likelihood score $\mathcal{P}_y(\vec{x}^*)$.

A further advantage of this probabilistic formulation is explicit out-of-distribution (OOD) detection. 
\add{To maximize sensitivity, OOD detection is evaluated over the complete descriptor set $\mathcal{D} = \{1,\dots,M\} \supseteq \mathcal{D}_\text{cls}$ (see Automatic Feature Selection, above) rather than the $m$ descriptors used for classification alone.} 
\add{For each descriptor $d \in \mathcal{D}$ and class $y$, we construct $p_{d,y}(x)$ following the same KDE procedure described above, with bandwidth $\sigma_d$ selected independently via the same five-fold cross-validated grid search.
We define an environment as an outlier if $p_{d,y}(x) < P_\text{cut}$ for all classes $y$, for any descriptor $d \in \mathcal{D}$.}
This criterion ensures that the vast majority of in-distribution configurations are retained while enabling detection of novel or anomalous environments, a capability essential for analyzing structural transformations. 
\add{In this work, we choose $P_\text{cut} = 5 \times 10^{-4}$. 
We find that our results are robust up to $P_\text{cut} = 1 \times 10^{3}$, beyond which outlier configurations are observed during compression/decompression trajectories.
This threshold can be adjusted to make the classifier more or less conservative, depending on the application requirements.}

\subsection{Training Data and Simulation Details}

All configurations were obtained from the Deep Potential (DP) MD trajectories reported in Ref.~\citenum{Szukalo:2025}.  
Two machine-learned potentials were employed, providing complementary representations of water's potential energy surface.  
The first, DP\_SCAN, was trained on electronic-structure data generated with the SCAN density functional~\cite{Sun:2015, Zhang:2021pd}, a meta-GGA functional known to reproduce many of water's structural and thermodynamic properties.  
The second, DP\_MBpol, was trained on the many-body polarizable (MBpol) potential~\cite{Babin:2013, Babin:2014, Medders:2014, Bore:2023}, which is parameterized against coupled-cluster reference data and is among the most accurate models of water available.  
All simulations were performed using LAMMPS (2~Aug~2023)~\cite{Thompson:2022} interfaced with DeePMD-kit (v3.0.0) \cite{Zhang:2018, Wang:2018}, \add{with a timestep of 0.5~fs. 
Simulations employed a system of 512 water molecules in a cubic box with periodic boundary conditions and the temperature and pressure were controlled via a Nos\'{e}--Hoover thermostat and barostat, respectively.}

\add{LDA configurations were generated by isobaric quenching from equilibrated liquid water configurations at constant cooling rates of $q_c = 1$~K/ns and $q_c = 10$~K/ns, at 1~bar.
Isothermal compression and decompression simulations were conducted starting from LDA configurations at constant rates of 
$q_p = 300$~bar/ns or $q_p = 3000$~bar/ns.
For each pressure and cooling rate, at least three independent simulations were performed to assess statistical variance.}

The training and evaluation dataset comprises of 10,000 molecular environments, equally divided between LDA and HDA.
Configurations were selected from thermodynamic conditions where phase identity is unambiguous, avoiding the transition region to ensure reliable labels. LDA environments were drawn from isobaric quenches at temperatures at least 30 K below the glass transition and from isothermal compression at pressures at least 3 kbar below the transformation pressure to HDA. 
HDA environments were taken from compression trajectories at least 3 kbar above the LDA-to-HDA transformation pressure and from decompression trajectories, the latter restricted to densities exceeding that of recovered HDA at 80 K and 1~bar to exclude partially transformed structures. 
As a benchmark for OOD detection, we curated 1,000 hexagonal ice (Ih) configurations. 
Ice Ih shares the tetrahedral hydrogen-bond network of LDA but exhibits long-range crystalline order absent in either amorphous phase, making it well-suited for evaluating OOD detection. 

\section{Benchmarking Model Performance for DP\_SCAN Environments}
\label{sec:S2}

Martelli and co-workers introduced a supervised neural network (BOO-NN) trained exclusively on bond-orientational order parameters, learning nonlinear mappings from BOO features to phase labels and adopting the traditional notion that orientational symmetry is the key structural discriminator between amorphous phases~\cite{Martelli:2020, Faure:2024}.  
DeFever \textit{et al.} proposed a conceptually different strategy with PointNet~\cite{Defever:2019}, which bypasses predefined descriptors altogether by learning symmetry-invariant representations directly from atomic coordinates using a permutation-invariant architecture; although highly successful for classifying crystallization and melting pathways, PointNet has not previously been applied to amorphous ices.  
Boattini and co-workers developed an unsupervised autoencoder-based model~\cite{Boattini:2019} that compresses BOO descriptors into a low-dimensional latent space before identifying structural motifs via Gaussian mixture modeling.  
To enable a supervised comparison, we adapt this method (AE–GMM) by fitting class-conditional multivariate Gaussians to the learned embeddings, capturing nonlinear correlations among BOO features while introducing a latent representation that is less directly interpretable.  
Collectively, these baselines cover descriptor-based neural networks, coordinate-level representation learning, and latent-space probabilistic modeling, offering a comprehensive set of benchmarks for contextualizing the strengths of our probabilistic classifier.

Here, we provide precisions, recalls, and accuracies of BOO-NN, PointNet, AE-GMM, and the present work on LDA/HDA classification for local atomic environments generated by DP\_MBpol (Table \ref{table:mbpol}) and DP\_SCAN (Table \ref{table:scan}).

\begin{table*}[htb]
\captionsetup{justification=raggedright, singlelinecheck=false,font=small}
\caption{Comparison of model performance for classification of DP\_MBpol-generated LDA/HDA environments. Values represent the mean and standard deviation over five-fold cross-validation.}
\centering
\begin{tabular}{cccccc}
\toprule
Model & HDA-Precision & HDA-Recall & LDA-Precision & LDA-Recall & Accuracy \\
\midrule
BOO-NN & 0.997 $\pm$ 0.001 & 0.997 $\pm$ 0.002 & 0.997 $\pm$ 0.002 & 0.997 $\pm$ 0.001 & 0.997 $\pm$ 0.001 \\
PointNet & 0.937 $\pm$ 0.008 & 0.988 $\pm$ 0.003 & 0.988 $\pm$ 0.003 & 0.934 $\pm$ 0.008 & 0.961 $\pm$ 0.003 \\
AE-GMM & 0.938 $\pm$ 0.008 & 0.982 $\pm$ 0.003 & 0.983 $\pm$ 0.003 & 0.935 $\pm$ 0.008 & 0.959 $\pm$ 0.003 \\
\midrule
Present Work (BOO) & 0.997 $\pm$ 0.001 & 0.968 $\pm$ 0.010 & 0.967 $\pm$ 0.010 & 0.997 $\pm$ 0.001 & 0.982 $\pm$ 0.005 \\
Present Work (ACSF) & 0.973 $\pm$ 0.012 & 0.988 $\pm$ 0.009 & 0.987 $\pm$ 0.010 & 0.972 $\pm$ 0.013 & 0.980 $\pm$ 0.003 \\
Present Work (BOO+ACSF) & 0.996 $\pm$ 0.001 & 0.989 $\pm$ 0.003 & 0.989 $\pm$ 0.003 & 0.996 $\pm$ 0.001 & 0.993 $\pm$ 0.001 \\
\bottomrule
\end{tabular}
\label{table:mbpol}
\end{table*}

\begin{table*}[htb]
\captionsetup{justification=raggedright, singlelinecheck=false,font=small}
\caption{Comparison of model performance for classification of DP\_SCAN-generated LDA/HDA environments. Values represent the mean and standard deviation over five-fold cross-validation.}
\centering
\begin{tabular}{cccccc}
\toprule
Model & HDA-Precision & HDA-Recall & LDA-Precision & LDA-Recall & Accuracy \\
\midrule
BOO-NN & 0.947 $\pm$ 0.007 & 0.997 $\pm$ 0.001 & 0.997 $\pm$ 0.001 & 0.945 $\pm$ 0.007 & 0.971 $\pm$ 0.003 \\
PointNet & 0.937 $\pm$ 0.008 & 0.988 $\pm$ 0.003 & 0.988 $\pm$ 0.003 & 0.934 $\pm$ 0.008 & 0.961 $\pm$ 0.003 \\
AE-GMM & 0.938 $\pm$ 0.008 & 0.982 $\pm$ 0.003 & 0.983 $\pm$ 0.003 & 0.935 $\pm$ 0.008 & 0.959 $\pm$ 0.003 \\
\midrule
Present Work (BOO) & 0.951 $\pm$ 0.006 & 0.977 $\pm$ 0.008 & 0.978 $\pm$ 0.008 & 0.950 $\pm$ 0.005 & 0.964 $\pm$ 0.006 \\
Present Work (ACSF) & 0.935 $\pm$ 0.010 & 0.997 $\pm$ 0.001 & 0.998 $\pm$ 0.001 & 0.931 $\pm$ 0.010 & 0.964 $\pm$ 0.005 \\
Present Work (BOO+ACSF) & 0.948 $\pm$ 0.006 & 0.987 $\pm$ 0.004 & 0.988 $\pm$ 0.004 & 0.945 $\pm$ 0.005 & 0.967 $\pm$ 0.004 \\
\bottomrule
\end{tabular}
\label{table:scan}
\end{table*}

Performance metrics are lower for classifying environments generated by DP\_SCAN than they are for DP\_MBpol. 
Table~\ref{table:mbpol} shows that accuracies for some models are greater than $0.99$ for DP\_MBpol environments, while here all accuracies are roughly between $0.96$-$0.97$ for DP\_SCAN environments. 
The relative performance of the models, however, is similar; BOO-NN and our work perform best, followed by PointNet and AE-GMM. 
These results agree with what is observed for DP\_MBpol environments: our work maintains comparable accuracy to deep learning methods while adopting a physically interpretable classification scheme. 
Interestingly, BOO-NN, AE-GMM, and our work show higher HDA recall and LDA precision than HDA precision and LDA recall. 
This indicates that these models are liberal in their assignment of HDA but conservative in their assignment of LDA, suggesting a relationship between HDA and LDA structures for DP\_SCAN not present in DP\_MBpol.

\newpage
\section{Out-of-distribution Predictions for DP\_SCAN and DP\_MBpol Environments}
\label{sec:S3}

In the main text, Section 2.1. we analyze the behavior of BOO-NN, PointNet, AE-GMM, and this work when making predictions on out-of-distribution atomic environments. 
Specifically, models are trained on HDA and LDA atomic environments, but are used to predict the labels of hexagonal ice (Ih) atomic environments.
In the main text, we only perform this analysis for environments generated by DP\_MBpol. 
Here, we report the same results for DP\_SCAN configurations.

\begin{figure*}[h]
    \centering
    \includegraphics[width=0.5\textwidth]{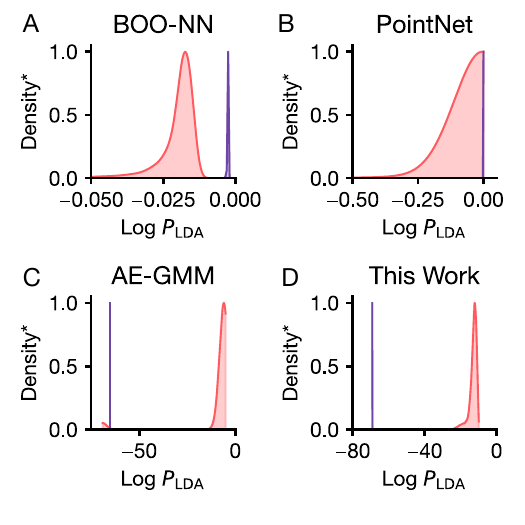}
    \captionsetup{justification=raggedright, singlelinecheck=false,font=small}
    \caption{Predicted LDA log probabilities from (A) BOO-NN, (B) PointNet, (C) AE-GMM, and (D) this work applied to DP\_SCAN-generated environments of LDA and hexagonal ice (Ih). Distributions for true LDA configurations are shown in red, while distributions for Ih configurations are shown in purple. Each density is normalized so that its maximum value is 1.0.}
    \label{fig:outlier_scan}
\end{figure*}

As observed in the main text, BOO-NN (Fig. \ref{fig:outlier_scan}A) and PointNet (Fig. \ref{fig:outlier_scan}B) struggle to separate LDA and Ih distributions, and these models confidently label Ih environments as LDA. 
Accordingly, the present work is capable of accurately distinguishing Ih environments as outliers from those on which the model was trained (\textit{i.e.}, LDA and HDA), as shown by the separation between the distributions in Figure \ref{fig:outlier_scan}D.  
Interestingly, while AE-GMM (Fig. \ref{fig:outlier_scan}C) was capable of detecting outliers for DP\_MBpol environments, this model struggles to do so for configurations generated by DP\_SCAN. 
Such a result demonstrates the utility of considering all available features, as is done in our work, relative to density estimation on reduced feature spaces, as is done by AE-GMM.

\newpage
\section{Out-of-distribution Prediction for Alternative Phases of Ice}
\label{sec:S4}

\add{
Our outlier detection method identifies local atomic environments that differ 
from the LDA/HDA training distribution along any structural dimension. 
To evaluate how the choice of features affects out-of-distribution (OOD) accuracy, we compare 
two approaches applied to an expanded dataset of ice polymorphs obtained from the datasets which were used in the training of the DP\_SCAN and DP\_MBpol models~\cite{Zhang:2021pd, Bore:2023}}

\add{
The first approach uses only the five selected features most informative for LDA/HDA classification.
As discussed in the main text, this results in a sparse, interpretable model whose features directly reflect what distinguishes the two amorphous states from one another.
The second approach includes all available features after removing pairwise-correlated descriptors (Pearson $|r| > 0.8$), retaining a diverse, uncorrelated subset of the full 120-dimensional descriptor space.
This is not equivalent to using all 120 descriptors simultaneously.
Including strongly correlated features in a naive Bayes product model would double-count shared structural information and degrade OOD sensitivity.}

\begin{table*}[htb]
\captionsetup{justification=raggedright, singlelinecheck=false,font=small,font={color=black},labelfont={color=black}}
%\color{red}
\caption{Accuracy of classifiers that use different numbers of features for OOD classification. 
Models are used to classify structures of different ice phases generated by the DP\_MBpol and DP\_SCAN water models. 
Models can either use five features relevant to LDA/HDA classification ($n=5$) or use all uncorrelated provided structural descriptors ($n=120$).}
\centering
\begin{tabular}{c|cc|cc}
\toprule
\multicolumn{1}{c|}{Phase} & \multicolumn{2}{c|}{DP\_MBpol} & \multicolumn{2}{c}{DP\_SCAN} \\
 & $n=5$ & $n=120$ & $n=5$ & $n=120$ \\
\midrule
Ice Ih & 1.000 & 1.000 & 1.000 & 1.000 \\
Ice Ic & 1.000 & 1.000 & 0.998 & 0.998 \\ 
Ice II & 0.000 & 0.122 & 0.003 & 0.378 \\ 
Ice III & 0.009 & 0.933 & 0.001 & 1.000 \\ 
Ice IV & 0.012 & 0.033 & 0.048 & 0.071 \\ 
Ice V & 0.003 & 0.872 & 0.064 & 0.998 \\ 
Ice VI & 0.024 & 0.134 & 0.087 & 0.250 \\ 
Ice VII & 1.000 & 1.000 & 1.000 & 1.000 \\ 
Ice VIII & 1.000 & 1.000 & 1.000 & 1.000 \\ 
Ice IX & 0.005 & 0.993 & 0.000 & 1.000 \\ 
Ice XI & 0.993 & 0.998 & 1.000 & 1.000 \\ 
Ice XII & 0.062 & 0.991 & 0.30 & 1.000 \\ 
Ice XIII & 0.006 & 0.763 & 0.052 & 0.980 \\ 
Ice XIV & 0.031 & 0.987 & 0.312 & 1.000 \\ 
Ice XV & 0.061 & 0.214 & 0.264 & 0.470 \\ 
\bottomrule
\end{tabular}
\label{table:ice_ood}
\end{table*}

\add{Table~\ref{table:ice_ood} shows OOD detection rates for both approaches across 14 ice polymorphs. 
Using the full uncorrelated descriptor set outperforms the five-feature model in most cases, with several phases moving from near-zero to near-complete detection. 
This indicates that the combined ACSF and BOO descriptor set contains structural information sufficient to distinguish the majority of ice polymorphs from amorphous ice. 
A subset of high-pressure phases (e.g., Ice II, IV, VI, and XV) remain only partially distinguishable from LDA/HDA even when using the full descriptor set.}

\add{This residual ambiguity is a consequence of the descriptor selection procedure used here, rather than the local environment representation itself.
At the level of individual atomic neighborhoods, some high-pressure ice phases share geometric features with LDA/HDA that are not fully resolved by the descriptors identified that optimally distinguish LDA and HDA. 
This does not indicate that these crystalline phases are physically present within computational LDA or HDA.
Independent molecular dynamics studies of LDA and HDA have found no evidence of ice-like domains in either amorphous phase~\cite{Martelli:2018}. 
Taken together, these results suggest that the observed classifier ambiguity for certain high-pressure ices arises from local geometric similarity between amorphous and crystalline packing motifs along the LDA/HDA classification axis, rather than from partially crystallized regions embedded within the amorphous structures. 
Distinguishing such phases would be possible if a classifier were directly trained to resolve the amorphous states from the crystalline ones.}

\newpage
\section{Sensitivity of Structural Pathways to Selected Feature Spaces}
\label{sec:S5}

In the main text, Section 2.6 analyzes how the evolution of average atomic environments proceeds during compressions and decompressions across a range of trajectories at different temperatures. 
We show these evolutions for DP\_MBpol and DP\_SCAN for two different feature spaces. 
Here, we show these evolutions for 25 feature spaces comprised of each pair of features selected by our model to be most informative for accurate LDA/HDA classification. 

\begin{figure*}[h]
    \centering
    \includegraphics[width=0.9\textwidth]{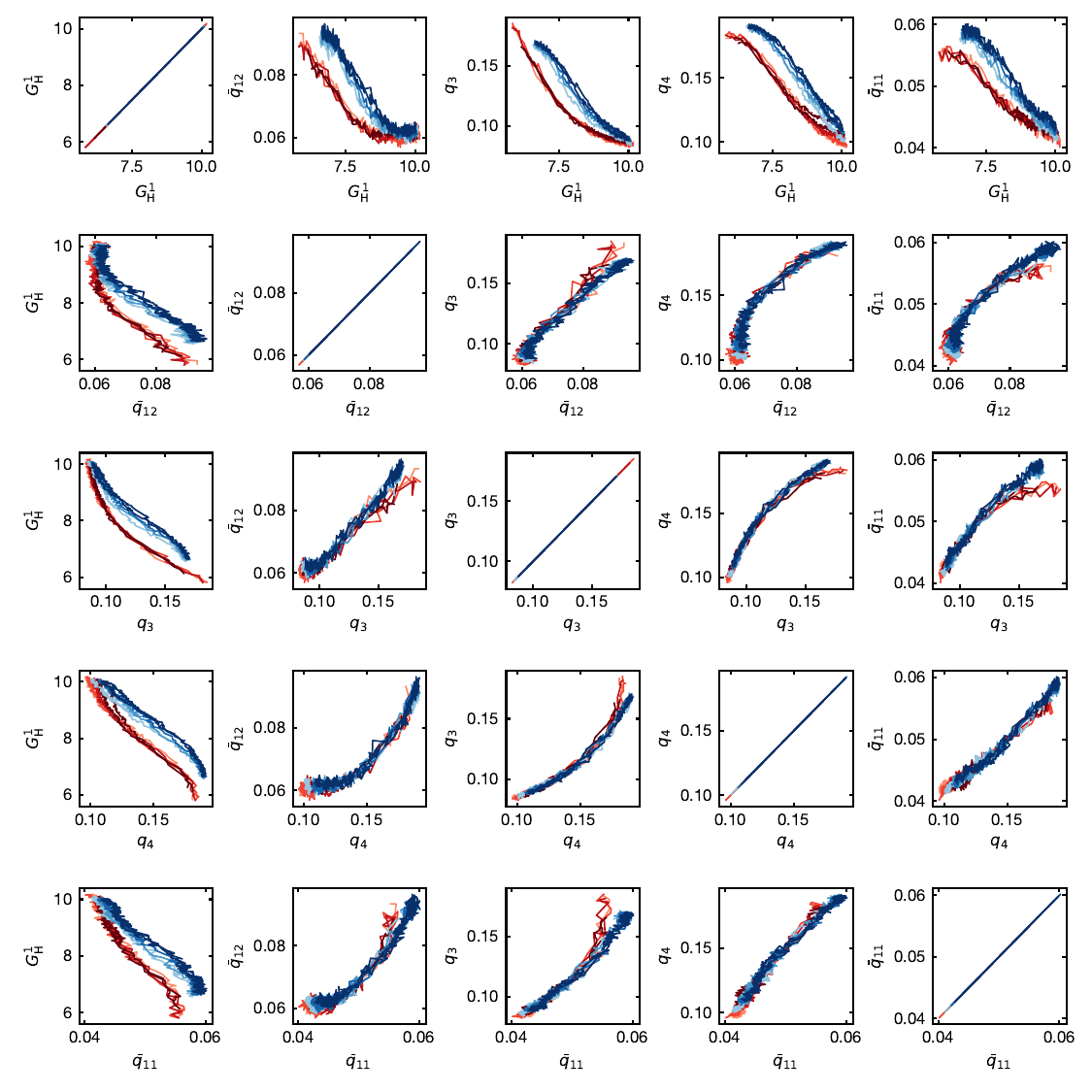}
    \captionsetup{justification=raggedright, singlelinecheck=false,font=small}
    \caption{Evolution of average local environment descriptors during compressions (blue) and decompressions (red) at temperatures from 60~K to 140~K for the DP\_MBpol force field. Descriptors are plotted in feature spaces for every pair of features selected by our model for LDA/HDA classification.
    Darker red/blue shades correspond to lower temperatures.}
    \label{fig:grid_mbpol}
\end{figure*}

\begin{figure*}[h]
    \centering
    \includegraphics[width=0.9\textwidth]{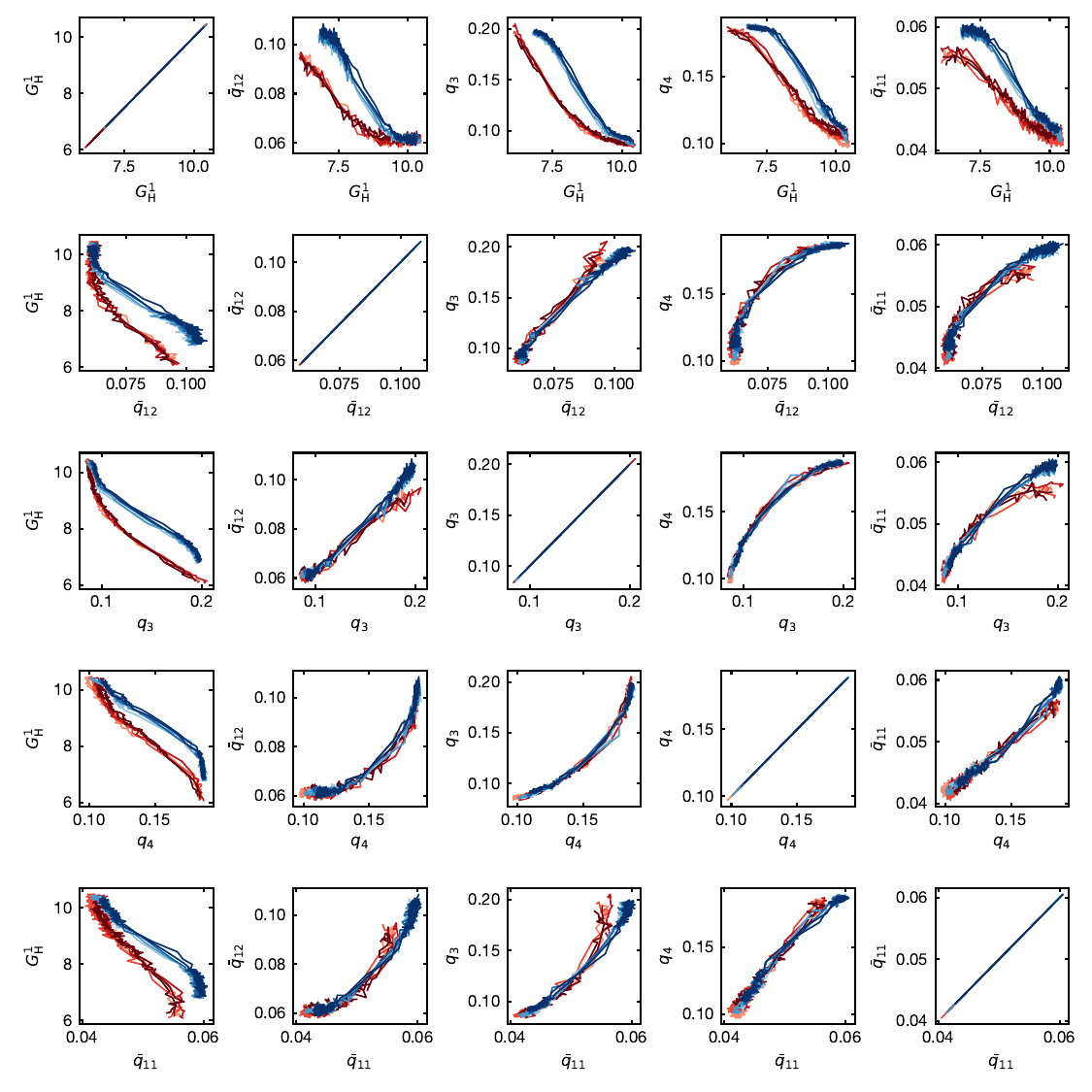}
    \captionsetup{justification=raggedright, singlelinecheck=false,font=small}
    \caption{Evolution of average local environment descriptors during compressions (blue) and decompressions (red) at temperatures from 60~K to 140~K for the DP\_SCAN force field. Descriptors are plotted in feature spaces for every pair of features selected by our model for LDA/HDA classification.
    Darker red/blue shades correspond to lower temperatures.}
    \label{fig:grid_scan}
\end{figure*}

Results from Figures \ref{fig:grid_mbpol} and \ref{fig:grid_scan} corroborate the results discussed in Section 2.6. 
Namely, we observe a hysteresis in environment pathways when considering feature spaces that consist of an ACSF and BOO, but this separation is not visible when visualizing pairs of BOOs. 
Therefore, it is necessary to consider both radial and spherical information to identify differences in the structural pathways taken by compressions and decompressions---an observation that would go unnoticed if we relied solely on the BOOs employed in previous studies.

\newpage
\clearpage
\section{Differences between DP\_SCAN- and DP\_MBpol-generated HDA Environments}
\label{sec:S6}

\begin{figure*}[h]
    \centering
    \includegraphics[width=0.5\textwidth]{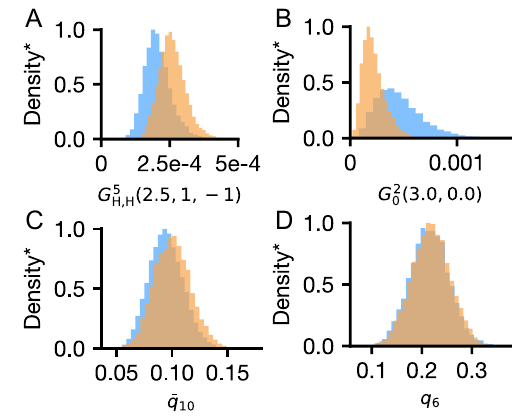}
    \captionsetup{justification=raggedright, singlelinecheck=false,font=small}
    \caption{Differences between local environments of HDA generated from DP\_SCAN and DP\_MBpol. 
    (A--D) Distributions of select ACSF and BOO descriptors for HDA generated by each model. 
    All probability densities are normalized such that the maximum value of either distribution is 1.0 (denoted as Density$^*$).
    The blue and orange colors represent the DP\_SCAN and DP\_MBpol models, respectively.}
    \label{fig:force_field_hda}
\end{figure*}

In the main text, Section 2.7 analyzes structural differences between LDA environments generated by DP\_SCAN and DP\_MBpol. 
Here, we consider structural differences between HDA environments generated by the same two force fields. 
Our classifier can distinguish between DP\_SCAN and DP\_MBpol HDA environments with an accuracy of $64.3 \pm 0.03$\%, indicating that there are structural differences (albeit less than LDA) present in these environments. 
We consider how distributions of DP\_SCAN and DP\_MBpol environments vary along a select set of structural descriptors with high mutual information for DP\_SCAN/DP\_MBpol classification, which we visualize in Figure \ref{fig:force_field_hda}.

Figures~\ref{fig:force_field_hda}A and~\ref{fig:force_field_hda}B show distributions of DP\_SCAN and DP\_MBpol environments along the two most informative features identified by our classifier. 
One can clearly observe differences in the two distributions, suggesting that (similar to LDA environments) DP\_SCAN and DP\_MBpol disagree most in their reproduction of interatomic distances. 
This is most directly observed when considering the second descriptor, $G_\text{O}^{2}(3.0, 0.0)$, where higher values indicate reduced oxygen--oxygen pairwise distances. 
Figures~\ref{fig:force_field_hda}C and~\ref{fig:force_field_hda}D show distributions of DP\_SCAN and DP\_MBpol environments along the two most informative BOOs. 
HDA environments appear to have similar distributions along these descriptors, indicating that DP\_SCAN and DP\_MBpol predict similar spherical symmetry in HDA environments.

\end{document}